\documentclass[prl,showpacs,twocolumn,amsmath,amssymb,superscriptaddress]{revtex4-1}
\usepackage{graphicx}
\usepackage{dcolumn}
\usepackage{bm}

\usepackage{graphicx}
\usepackage{dcolumn}
\usepackage{bm}

\usepackage{amsmath,amssymb}

\usepackage{color}

\newcommand{\pbar}{\bar{p}}
\newcommand{\bS}{{\bf S}}

\newcommand{\la}{\langle}          
\newcommand{\ra}{\rangle}
\newcommand{\bea}{\begin{eqnarray}}                    
\newcommand{\eea}{\end{eqnarray}}

\begin{document}

\title{Deconfined criticality in the frustrated Heisenberg honeycomb antiferromagnet}

\author{R. Ganesh}
\affiliation{Institute for Theoretical Solid Sate Physics, IFW Dresden, Helmholtzstr. 20, 01069 Dresden, Germany}

\author{Jeroen van den Brink}
\affiliation{Institute for Theoretical Solid Sate Physics, IFW Dresden, Helmholtzstr. 20, 01069 Dresden, Germany}
\affiliation{Department of Physics, Technical University Dresden, D-1062 Dresden, Germany}

\author{Satoshi Nishimoto}
\affiliation{Institute for Theoretical Solid Sate Physics, IFW Dresden, Helmholtzstr. 20, 01069 Dresden, Germany}

\date{\today}

\begin{abstract}
Using the density-matrix renormalization group, we determine the phase diagram of the spin 1/2 Heisenberg antiferromagnet on a honeycomb lattice with a nearest neighbor interaction $J_1$ and a frustrating, next-neighbor exchange $J_2$. As frustration increases, the ground state exhibits N\'eel, plaquette and dimer orders, with critical points at $J_2/J_1=0.22$ and $0.35$.
We observe that 
both the spin gap and the corresponding order parameters vanish {\it continuously} at {\it both} the critical points, indicating the presence of deconfined quantum criticality.
\end{abstract}

\maketitle

\paragraph{Introduction}
Models of frustrated magnetism on the honeycomb lattice have lately received tremendous interest. This interest stems from sign-problem-free Quantum Monte Carlo (QMC) studies which have established the presence of a spin liquid phase in the honeycomb Hubbard model~\cite{Meng10}. Approaching from the strong coupling side, the physics at intermediate values of the Hubbard interaction $U$, for which the novel spin liquid phase has been found, can be described by the spin 1/2 Heisenberg model characterized by an antiferromagnetic interaction $J_1$ between neighboring spins and a frustrating, next-nearest neighbor exchange $J_2$. When the frustration is small and $J_2$ weak, the well-known N\'eel ordered state is stable, but at a critical value of $\alpha = J_2/J_1$ it gives way to another, possibly liquid, phase. 
While all studies so far agree upon the presence of a phase transition, the nature of this intermediate phase that is reached by the transition out of the N\'eel  state is heavily debated.  The intermediate phase has been identified as a $Z_2$ spin liquid by some~\cite{Wang10,Lu11,Clark11} and as a plaquette-Resonating Valence Bond (pRVB) state, breaking translational symmetry, by others~\cite{Mosadeq11,Albuquerque11,Bishop12}. A recent variational calculation argues instead that the intermediate state does not have plaquette order~\cite{Mezzacapo12}. Upon further increasing the frustration parameter $\alpha$, a second transition takes place into a ground state that breaks lattice rotational symmetry but may or may not have magnetic order. 

We analyze this complex situation by formulating and answering four succinct 
fundamental questions on the $J_1-J_2$ honeycomb Heisenberg model: (i) 
As to the N\'eel state: do quantum fluctuations tend to stabilize or destroy 
it? In other words, does N\'eel order vanish above or below the classical 
threshold of $\alpha = 1/6$? (ii) What is the nature of the intermediate state?
 Is it a liquid state or does it have plaquette order? (iii) What is the ground 
state for large $\alpha$? Does it have magnetic order? (iv) What is the nature of the two phase transitions? Do the order parameters develop discontinuously or 
continuously across the quantum critical points?

\begin{figure}
 \includegraphics[width=.8\columnwidth]{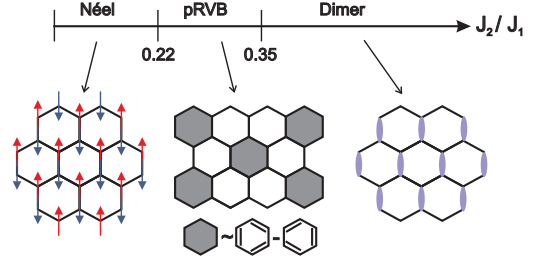}
\caption{Phase diagram of the spin 1/2 Heisenberg antiferromagnet on a honeycomb lattice with a nearest neighbor interaction $J_1$ and a frustrating, next-neighbor exchange $J_2$ as obtained from DMRG.}
\label{fig.phasediagram}
\end{figure}

We use nominally-exact two-dimensional density-matrix renormalization group 
(DMRG) calculations to settle these issues and establish that: (i) N\'eel
 order is stabilized beyond the classical limit, up to $\alpha_{c1}=0.22$ (ii) the intermediate state has weak plaquette order with $f$-wave symmetry, 
and (iii) for $\alpha_{c2} > 0.35$, the ground state has dimer order 
and breaks lattice rotational symmetry. These results are summarized in the 
phase diagram shown in Fig.~\ref{fig.phasediagram}. Moreover, we find that within numerical precision,  (iv) both the spin gap and the relevant order 
parameters vanish continuously,  at both critical points $\alpha_{c1}$ and 
 $\alpha_{c2}$. This implies that even if two different symmetries are 
broken on either side of $\alpha_{c}$, the transition is {\it not} first-order, as one would expect from a Ginzburg-Landau-type theory.  Having two second-order transitions between the N\'eel, plaquette and dimer phases, implies that the critical theory for these transitions is unusual and is not described in terms of the order parameter fields of either phase. It indicates instead the presence of two deconfined quantum critical points~\cite{Senthil04,Xu12}.

\paragraph{Frustrated honeycomb Heisenberg model}
The Hamiltonian corresponding to the $J_1-J_2$ Heisenberg model on a  honeycomb lattice is 
\bea
H = J_1 \sum_{\la ij \ra} \bS_{i} \cdot \bS_{j} + J_2 \sum_{\la \la ij \ra\ra} \bS_{i} \cdot \bS_{j},
\label{eq.Hmlt}
\eea
where $\la ij \ra$ and  $\la \la ij \ra \ra$ denote nearest neighbor and next-neighbor sites $i$ and $j$, respectively, and  $\alpha= J_2/J_1$ parameterizes 
the strength of the frustration. We consider antiferromagnetic coupling:
 $J_1$, $J_2$ and $\alpha$ are all positive. The model is well understood 
in the classical limit: at the critical value of  $\alpha=1/6$, N\'eel order 
gives way to a spiral state with interesting order-by-disorder 
physics~\cite{Mulder10,Okumura10}. However, in the extreme quantum limit 
of $S=1/2$, the phase diagram is not well established~\cite{Wang10,Lu11,Clark11,Mosadeq11,Albuquerque11,Bishop12,Mezzacapo12,Clark11}. We use DMRG to resolve this issue. 



\paragraph{Method}
Our DMRG is truly two-dimensional -- we consider clusters with various 
geometries chosen to be conducive to various ordering patterns. It is well 
known that one can lift the degeneracy of wave functions by taking some or all edges to be open. We use appropriate edge geometries as weak perturbing fields to induce symmetry breaking in the ground state. By performing measurements in the center of the cluster, one can estimate the order parameter induced by the edge geometry.  Upon systematically increasing the size of the system, the effect of the edges becomes progressively weaker and thus, by scaling to the thermodynamic limit, we can obtain the value of the order parameter in the ground state. In all cases, we have obtained smooth finite size scaling which indicates that our results exhibit a steady convergence to the thermodynamic limit.

As described below, we have used a variety of cluster geometries appropriate for each phase. Note that the performance of DMRG calculation is equally stable for any ordered phase at $\alpha<{\cal O}(1)$. We study several cluster sizes with total number of sites up to 96 and keep up to 6000 density-matrix eigenstates 
in the renormalization procedure. We perform $\sim 10$ sweeps until the 
ground-state energy converges within an error of $\sim 10 ^{-5} J_1$. All 
quantities calculated in this letter have been extrapolated to the limit $n \to \infty$, where $n$ is the number of retained eigenstates. 
 
\begin{figure}
\includegraphics[width=\columnwidth]{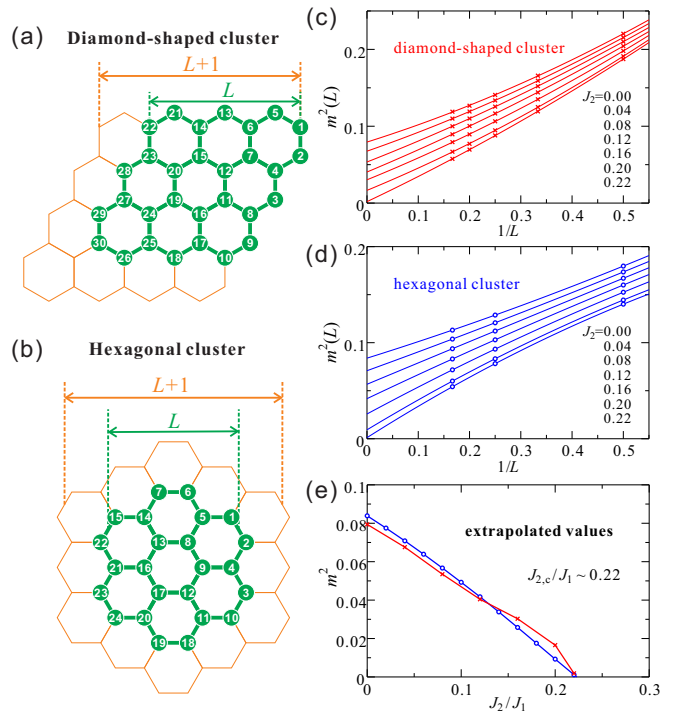}
\caption{Finite size scaling of N\'eel order parameter. (
a) Diamond cluster with $L=3$. (b) Hexagonal cluster with $L=3$. (c-d) 
Finite size scaling of N\'eel order paramater defined in Eq. \ref{eq.NeelOP} 
for diamond and hexagonal clusters. (e) Scaled N\'eel order parameter as a 
function of $\alpha=J_2/J_1$ for diamond (red, closed circles) and hexagonal 
(blue, open circles) clusters.} 
\label{fig.Neelscaling}
\end{figure}
 
\paragraph{Quantum stabilitization of N\'eel order}
We first determine the value of $\alpha$ at which N\'eel order vanishes and 
establish the role of quantum fluctuations in this process. Na\"ively, one 
expects quantum fluctuations to destabilize N\'eel order for $S=1/2$, thereby 
pushing the $\alpha_{c1}$ to a value below $1/6$. On the other hand, as the 
N\'eel state is collinear, quantum fluctuations may prefer the N\'eel state 
over a competing spiral phase and push $\alpha_{c1}$ above $1/6$. Even though
various approaches have been used to resolve this issue, a consistent picture 
has not emerged so far. Calculations which support the hypothesis that 
$\alpha_{c1}< 1/6$ include linear spin wave theory~\cite{Mattsson94}, 
one-loop renormalization group study of the non-linear sigma 
model~\cite{Einarsson91}, functional renormalization group 
analysis~\cite{Reuther11}, and a Variational Monte Carlo (VMC) approach using 
RVB and Huse-Elser wavefunctions~\cite{Clark11}. On the other hand, approaches which support the $\alpha_{c1} > 1/6$ hypothesis include exact diagonalization (ED)~\cite{Mosadeq11,Albuquerque11}, Schwinger boson mean-field 
theory~\cite{Mattsson94}, series expansions~\cite{Oitmaa11}, coupled-cluster
 calculations~\cite{Bishop12} and a VMC calculation using entangled plaquette states~\cite{Mezzacapo12}.

The DMRG results presented in Fig.~\ref{fig.Neelscaling} conclusively establish that quantum fluctuations stabilize N\'eel order beyond the classical regime of stability. We have used two cluster geometries -- diamond and hexagonal 
[Fig.~\ref{fig.Neelscaling}a,b]. One should be aware that periodic boundary 
conditions in some direction artificially enhance or diminish N\'eel correlations due to short range periodicity. This finite-size effect decays only slowly with increasing cluster size. To circumvent this issue, 
we keep all edges of the clusters open and measure the following order 
parameter as a function of $\alpha$:
\bea
m^2(N) = \frac{1}{N}\left( \sum_i (-1)^i \vec{S}_i \right)^2.
\label{eq.NeelOP}
\eea
As shown in Fig.\ref{fig.Neelscaling}, this quantity shows good finite-size 
scaling with terms proportional to $1/L$ and $1/L^2$, where $L$ is the linear 
extent of the system. In the unfrustrated situation ($\alpha=0$), the staggered moment $m$ in the thermodynamic limit comes out to be $0.2857 \pm 0.039$ which 
is consistent with previously estimated values of 0.2677(6) and 0.270
 obtained from QMC~\cite{Castro06} and ED~\cite{Albuquerque11} respectively. As $\alpha$ increases, the obtained value of the N\'eel order parameter steadily 
decreases. At the critical value of $\alpha_{c1}\sim 0.22$, we observe that N\'eel order vanishes in a continuous transition as shown in Fig.\ref{fig.Neelscaling}e. Both diamond and hexagonal cluster geometries give the same value of $\alpha_{c1}$, which signals the robustness of our result. Thus, quantum fluctuations stabilize N\'eel order significantly beyond the classical threshold. 

\paragraph{Non-linear spin wave analysis}
The excitations of the N\'eel state are well captured by spin wave theory, which treats quantum fluctuations using an expansion in powers of $S$. Linear spin wave theory with $\mathcal{O}(S^1)$ terms gives $\alpha_{c1} \sim 0.11$~\cite{Mattsson94}, which is {\it below} the classical threshold. To reconcile this with the observed DMRG phase boundary, we take into account the quartic spin wave interaction terms of order $\mathcal{O}(S^0)$. We treat the interactions at mean-field level (for details, see Supplementary Material) and observe that the 
Hartree Fock parameters merely renormalize the strength of the $J_1$ and $J_2$ couplings. This effectively scales the frustration parameter $\alpha=J_2/J_1$ down so that the N\'eel state only becomes unstable beyond $\alpha \sim 0.214$.  
The quartic terms thereby provide a significant correction to the critical frustration ratio. The precise value of $\alpha_{c1}$ may depend upon further corrections beyond quartic order. Nevertheless, non-linear
 spin wave analysis confirms the strong tendency for quantum fluctuations to stabilize N\'eel order beyond the classical limit.

\begin{figure}
\includegraphics[width=\columnwidth]{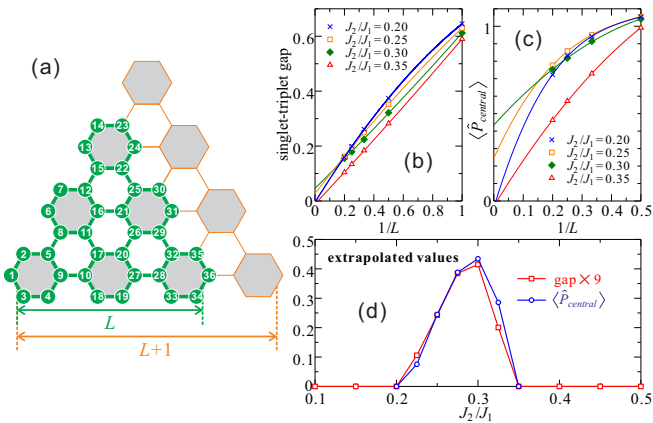}
\caption{(a): Cluster geometry used to establish the presence of plaquette order. (b,c) Finite size scaling of spin gap and $\la \hat{P}_{central}\ra$ -- a measure of pRVB amplitude. (d) Spin gap and $\la \hat{P}_{central}\ra$ in the thermodynamic limit.} 
\label{fig.plaquetteplots}
\end{figure}

\paragraph{Intermediate plaquette phase}
We observe the presence of an intermediate plaquette-RVB (pRVB) phase, as suggested previously~\cite{Fouet01, Mosadeq11, Albuquerque11, Bishop12}, for $0.22 \lesssim \alpha \lesssim 0.35$. This state consists of a $\sqrt{3}\times\sqrt{3}$ arrangement of plaquettes as shown in Fig.~\ref{fig.phasediagram} -- each shaded plaquette is in an anti-symmetric combination of the two Kekul\'e singlet covers. 
To test for plaquette order in the ground state, we choose the cluster geometry shown in Fig.~\ref{fig.plaquetteplots}a which favors plaquette order (this also favors columnar dimer order~\cite{Xu11}, but we have explicitly checked that it order does not occur). This choice of boundary conditions acts as a weak field which induces plaquette ordering as shown by the shaded hexagons in Fig.~\ref{fig.plaquetteplots}a. 
To determine the pRVB order parameter, we first define the two single-plaquette states $\vert a \ra$ and $\vert b \ra$ -- the two Kekul\'e singlet covers of a single hexagon. The $f$-wave, antisymmetric, pRVB wavefunction is given by $\vert - \ra \sim \vert a \ra - \vert b\ra$, upto a normalization constant. The order parameter corresponding to pRVB order is the projection onto the antisymmetric wavefunction: $\hat{O}_{pRVB} = \vert - \ra \la - \vert $ acting on a shaded plaquette in Fig.~\ref{fig.plaquetteplots}a. We use the closely related plaquette-flip operator which flips the two Kekul\'e covers:
\bea
\hat{P} = -\vert a \ra \la b \vert -\vert b \ra \la a \vert.
\label{eq.Pdef}
\eea
If the plaquette is in the pure $\vert -\ra$ state, this operator has expectation value $5/4$ (details in Supplementary Material). For the case of $s$-wave pRVB order, this expectation value would be negative. 

To determine the strength of the pRVB ordering at the cluster center, we define $\la \hat{P}_{central} \ra$ as the average of $\la \hat{P} \ra$ over three plaquette-ordering hexagons at the center of the system. As seen from our cluster geometry in Fig.~\ref{fig.plaquetteplots}a, one cannot always identify a single central plaquette for a given $L$. But we can always identify a central triad of plaquettes. Finite size scaling of $\la \hat{P}_{central} \ra$ provides the strength of pRVB order in the limit of infinite system size. Consistent with $f$-wave pRVB order, this expectation value is positive for $0.22\lesssim \alpha \lesssim 0.35$. Fig.~\ref{fig.plaquetteplots}c shows the finite size scaling of $\la\hat{P}_{central}\ra$ which indeed scales to a positive value in thermodynamic limit. Also we find a finite spin gap that is consistent with $\sqrt{3}\times \sqrt{3}$ plaquette ordering.
We note, however, that strong quantum fluctuations reduce the amplitude of plaquette ordering: $\la\hat{P}_{central}^{N=\infty}\ra$ reaches a maximum value of $\sim 0.43$ compared to $5/4$ for the case of pure pRVB order.  The strength of pRVB order can also be characterized by $\pbar$, the amplitude of the projection onto the $\vert - \ra$ plaquette wavefunction as defined in Ref.~\cite{Ganesh12}. For decoupled hexagons in the regime $0 < \alpha < 0.5$, there is  perfect pRVB order with $\pbar=1$. Our DMRG results indicate that in the honeycomb $J_1-J_2$ model, pRVB order is strongly affected by quantum fluctuations and reduced to $\pbar\lesssim0.43$.  To confirm the existence of pRVB order, we also measure the spin gap which is by definition the energy difference between the first triplet excited state and the singlet ground state,
\bea
\Delta(L)=E_1(L)-E_0(L), \ \ \ \Delta=\lim_{L \to \infty}\Delta(L),
\eea
where $E_n(L)$ is the $n$-th eingenenergy ($n=0$ corresponds to the ground state) of the system size $L$. The scaling analysis of the finite-size data is shown in Fig.~\ref{fig.plaquetteplots}b, and the results extrapolated to the thermodynamic limit are plotted in Fig.~\ref{fig.plaquetteplots}d.  We observe that the gap is finite only in the region of positive $\la \hat{P}_{central} \ra$.

\begin{figure}
\includegraphics[width=\columnwidth]{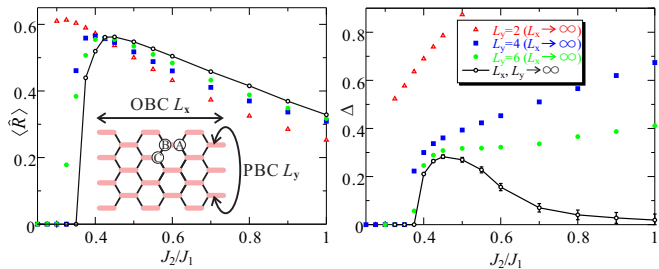}
\caption{Left: finite size scaling of $\la \hat R\ra$, order parameter corresponding to lattice rotational symmetry breaking. For finite size scaling, we first take $L_x \rightarrow \infty$ and then $L_y \rightarrow \infty$. We show data points obtained by $L_x$ scaling for different fixed $L_y$ values. The points connected by the line are final values obtained from $L_y$ scaling. Inset: cluster geometry to detect dimer order. We enforce periodic boundary conditions along $L_y$ and open boundary conditions along $L_y$. Right: The order parameter and spin gap (also obtained by similar finite size scaling) as a function of $\alpha=J_2/J_1$. }
\label{fig.dimerplots}
\end{figure}

\paragraph{Dimer phase}
At larger values of $J_2$, the presence of a dimer state which breaks lattice rotational symmetry has been proposed previously~\cite{Fouet01,Mulder10}. This state has been variously called the staggered-Valence Bond Solid (s-VBS) or the Nematic VBS state in literature. We establish that this state occurs in the phase diagram for $\alpha \gtrsim 0.35$ using the cluster geometry in Fig.\ref{fig.dimerplots}a. We use open boundary conditions in the $x$ direction and periodic boundary conditions along $y$, thus breaking the degeneracy associated with threefold lattice rotational symmetry. The cluster favors bond ordering with horizontal dimers as shown in Fig.\ref{fig.dimerplots}a.
We first measure the breaking of lattice rotational symmetry by evaluating the expectation value of
\bea
\hat{R} = \bS_{\rm A}\cdot \bS_{\rm B} - \bS_{\rm B}\cdot \bS_{\rm C},
\eea
where the sites A, B and C are chosen close to the center of the system (see Fig.~\ref{fig.dimerplots}a). We determine the expectation $\la \hat{R} \ra$ while systematically increasing system size. If the true ground state breaks lattice rotational symmetry, we expect this quantity to scale to a non-zero value in the thermodynamic limit. 
For finite size scaling, we first take $L_x \rightarrow \infty$ followed by $L_y\rightarrow \infty$. This sequence of limits ensures that there is no degeneracy arising from lattice rotations.  We obtain smooth finite size scaling by restricting ourselves to even values of $L_y$, as shown in Fig.\ref{fig.dimerplots}. Including odd $L_y$ values leads to small oscillations preventing smooth scaling. 

Fig.\ref{fig.dimerplots}b shows that $\la \hat{R} \ra$ scales to a non-zero value for $\alpha \gtrsim 0.35$, clearly establishing broken lattice rotational symmetry in the ground state. However, this is consistent with two ground state candidates -- dimer order or magnetic stripe order~\cite{Albuquerque11}. To distinguish between these two, we measure the spin gap. The finite size scaling for spin gap is shown in Fig.~\ref{fig.dimerplots}b. 
The error bars shown in Fig.~\ref{fig.dimerplots}b are associated with the choice of $\nu$ to fit the data points. For $0.35\lesssim \alpha \lesssim 0.6$, the spin gap scales to a non-zero value robustly. For $\alpha \gtrsim 0.7$,  it is not possible to determine reliably whether the spin gap closes. The non-zero spin gap clearly indicates dimer order and rules out the presence of broken spin rotation invariance. 

\paragraph{Nature of phase transitions}
We have clearly demonstrated the presence of N\'eel, plaquette and dimer orders. Na\"ively, one expects first-order quantum phase transitions (QPTs) between these phases as they break different symmetries. Our DMRG results, however,  evidence a continuous transition out of the N\'eel phase: as can be seen from Fig.~\ref{fig.Neelscaling}, the N\'eel order parameter vanishes continuously at $\alpha_{c1} = 0.22$. This implies the presence of an exotic deconfined QPT~\cite{Bishop12}. Approaching from the pRVB side, the quantum field theory governing this deconfined transition must involve spinons coupled to vortices in the pRVB order parameter~\cite{Xu11}. This is an exciting proposition as a deconfined QPT in a model with realistic Heisenberg interactions has not been identified before. Surprisingly, DMRG results suggest that also the plaquette-dimer transition is continuous. As seen from Fig.~\ref{fig.plaquetteplots}d and Fig.~\ref{fig.dimerplots}, at $\alpha_{c2} =0.35$, there is no evidence for either plaquette ordering or a breaking of lattice rotational symmetry. 
More detailed work will be needed to study the vicinity of these transitions, to extract critical exponents and to rule out weak first-order behavior or the presence  of a different small intervening phase. 
If there is indeed a continuous transition between dimer and plaquette phases, it would be yet another Landau-forbidden QPT within the same model. The field theory corresponding to this transition would be of immense interest.

\acknowledgments
We thank I. Rousochatzakis for many useful discussions.

Note added: during the preparation of this manuscript a DMRG study in Ref.~\cite{Zhu12} reported a similar sequence of  N\'eel, plaquette and dimer order as well as the continuous nature of the transition out of the N\'eel phase.

\bibliographystyle{apsrev}
\bibliography{Heisenberg_honeycomb}

\end{document}


\title{Supplementary Material for: \\ Deconfined criticality in the frustrated Heisenberg honeycomb antiferromagnet}

\author{R. Ganesh}
\affiliation{Institute for Theoretical Solid Sate Physics, IFW Dresden, Helmholtzstr. 20, 01069 Dresden, Germany}

\author{Jeroen van den Brink}
\affiliation{Institute for Theoretical Solid Sate Physics, IFW Dresden, Helmholtzstr. 20, 01069 Dresden, Germany}
\affiliation{Department of Physics, Technical University Dresden, D-1062 Dresden, Germany}

\author{Satoshi Nishimoto}
\affiliation{Institute for Theoretical Solid Sate Physics, IFW Dresden, Helmholtzstr. 20, 01069 Dresden, Germany}

\date{\today}

\maketitle 

\subsection{Non-linear spin wave theory}
\label{ssec.NLSWT}

To analyse the excitations about the N\'eel state, we follow the spin wave formalism of Ref.~\cite{Mulder}. Using the Holstein Primakoff representation, the Hamiltoniann 
is expanded in powers of the spin length $S$. Ultimately however, we will set $S=1/2$. The classical energy of the N\'eel state is given by terms proportional to $S^2$
\bea
E_{Cl} =NS^2 \Big[ -\frac{3}{2}J_1  + 3J_2 \Big].
\eea
where $N$ is the total number of spins. The quantum correction, of order $S$, is given by
\bea
H_{qu} = \sum_{k} \left(\begin{array}{cc} a_{\bk}^\dg & b_{-\bk} \end{array}\right)
\left(\begin{array}{cc}
A_{\bk} & B_{\bk}\\
B_{\bk}^* & A_{\bk}
      \end{array}\right)
\left(\begin{array}{c} a_{\bk} \\ b_{-\bk}^\dg  \end{array}\right),
\label{Eq.Hmlt}
\eea
where
\bea
\nn A_{\bk} &=& S\Big[ 3J_1 - 6J_2 + 2J_2\{ \cos k_a + \cos k_b + \cos(k_a+k_b)\}\Big], \\
\nn B_{\bk} &=& -S J_1  \gamma_{\bk}.
\eea
We have defined $\gamma_{\bk} = \sum_{\bdelta} e^{i \bk\cdot \bdelta}$, where $\bdelta$'s are the nearest neighbour vectors. The quantities $k_a$ and $k_b$ are components of momentum along two primitive lattice vectors of the triangular Bravais lattice. The operator $a_\bk^\dg$ ($b_\bk^\dg$) creates a spin wave excitation on the A (B) sublattice.
This Hamiltonian matrix can be diagonalized by a bosonic Bogoliubov transformation with the eigenvalue
\bea
\lambda_\bk = \sqrt{A_{\bk}^2 - \vert B_{\bk}\vert^2}.
\eea
For $J_2>J_1/6$, the spin wave energy $\lambda_{\bk}$ becomes complex near the $\Gamma$ point indicating that N\'eel order is unstable. We next include the quartic corrections arising from spin wave interactions by retaining terms of order $\mathcal{O}(S^{(0)})$. There are no cubic terms. The interaction terms proportional to $J_1$ are given by
\bea
\nn H_{J_1}(\mathcal{O}(S^{0})) = \frac{J_1}{4} \sum_{i,\bdelta} \left[
a_i b_{j}^\dg b_j b_j + a_i^\dg a_i a_i b_j \right. \\
\left. + a_i^\dg b_j^\dg b_j^\dg b_j + a_i^\dg a_i^\dg a_i b_j^\dg - 4 a_i^\dg a_i b_j^\dg b_j \right].
\eea
The index j stands for $i+\bdelta$. The quartic terms proportional to $J_2$ are given by
\bea
\nn H_{J_2}(\mathcal{O}(S^{0})) = \frac{-J_2}{8} \sum_{i,\boldeta} \Big[
a_i a_m^\dg a_m^\dg a_m \!\! + \! a_i^\dg a_i a_i a_m^\dg \\
\!+\! a_i^\dg a_m^\dg a_m a_m \!+\! a_i^\dg a_i^\dg a_i a_m
-4 a_i^\dg a_i a_m^\dg a_m \Big] + (a\rightarrow b).
\eea
The index m stands for $i+\boldeta$, where $\boldeta$ is a next-nearest neighbour vector. 

We treat these terms using the Hartree Fock approach. Using Wick's theorem, we replace bilinears with their expectation values and ignore the remaining normal ordered quartic interaction piece. We take only the following bilinears to have non-zero expectation values:
\bea
\la a_i^\dg a_i \ra &=& \la b_i^\dg b_i \ra = n,\\
\la a_i b_{i+\bdelta} \ra &=& p, \\
\la a_i^\dg a_{i+\boldeta} \ra  &=& h.
\eea                                                                                                                                                                                                                                                                                                                                                                                                          
These are the only bilinears which have non-zero expectation values within the quadratic theory. This choice of order parameters leads to a self-consistent theory that does not induce extra bilinears with non-zero values. The quantities $n$ and $h$, being expectation values of Hermitian operators, are real. We take $p$ to be real, as it is real within the quadratic theory. Using the symmetries of the underlying N\'eel state, we take these three quantitites to be independent of position and choice of neighbour ($\bdelta,\boldeta$).

The interaction terms can be decoupled as 
\bea
\nn H_{J_1}(\mathcal{O}(S^{0}))  &=& J_1 \sum_\bk \Big[ 3\{p-n\} ( a_{\bk}^\dg a_{\bk}              
+ b_{\bk}^\dg b_{\bk}) \\
\nn&+& \{ n-p \}\gamma_\bk (a_{-\bk} b_{\bk} + a_{\bk}^\dg b_{-\bk}^\dg )\Big].\\
\nn H_{J_2}(\mathcal{O}(S^{0})) &=& J_2 \sum_{\bk} {}\Big[ 6\{ n-h \} (a_{\bk}^\dg a_{\bk}) \\
&-& \{n-h\} \mu_{\bk} (a_{\bk}^\dg a_{\bk}) \Big]+(a\rightarrow b).
\eea
Here, $\mu_\bk = \sum_{\boldeta} \exp (i \bk \cdot \boldeta)$.
With this decoupling, these terms enter the quadratic Hamiltonian in Eq.\ref{Eq.Hmlt}. We have 
\bea
A \rightarrow A + 3 J_1 \{p-n\} - 6 J_2 \{h-n\} + J_2 \{h-n\} \mu_\bk  \nonumber
\eea
and
\bea
B \rightarrow B + J_1 \{ n-p \} \gamma_\bk. \nonumber
\eea

\begin{figure}
\includegraphics[width=2.75in]{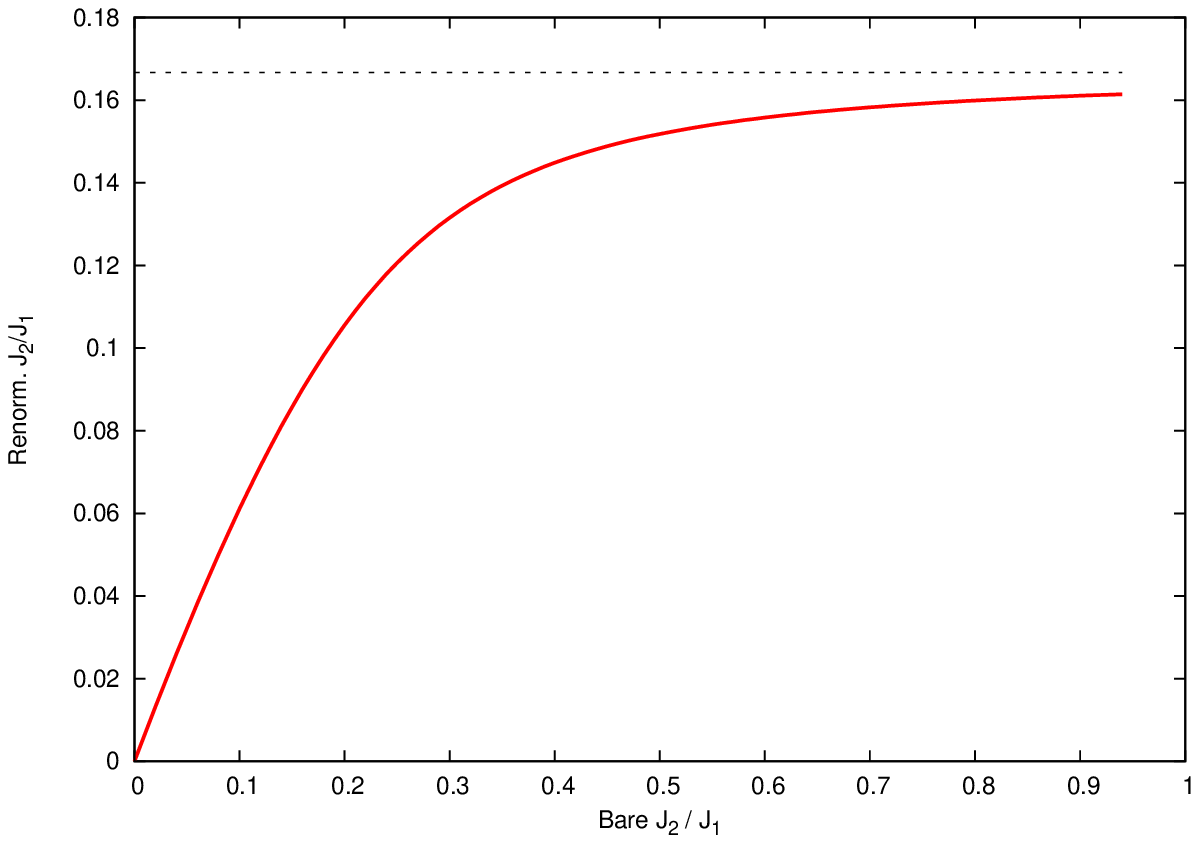}
\caption{Effective $J_2/J_1$ obtained after Hartree-Fock treatment of interactions. Note that the effective $J_2/J_1$ lies only approaches the instability threshold 1/6 when the bare ratio $\sim 0.95$. }
\label{Fig.effcoupling}
\end{figure}

\begin{figure}
\includegraphics[width=2.75in]{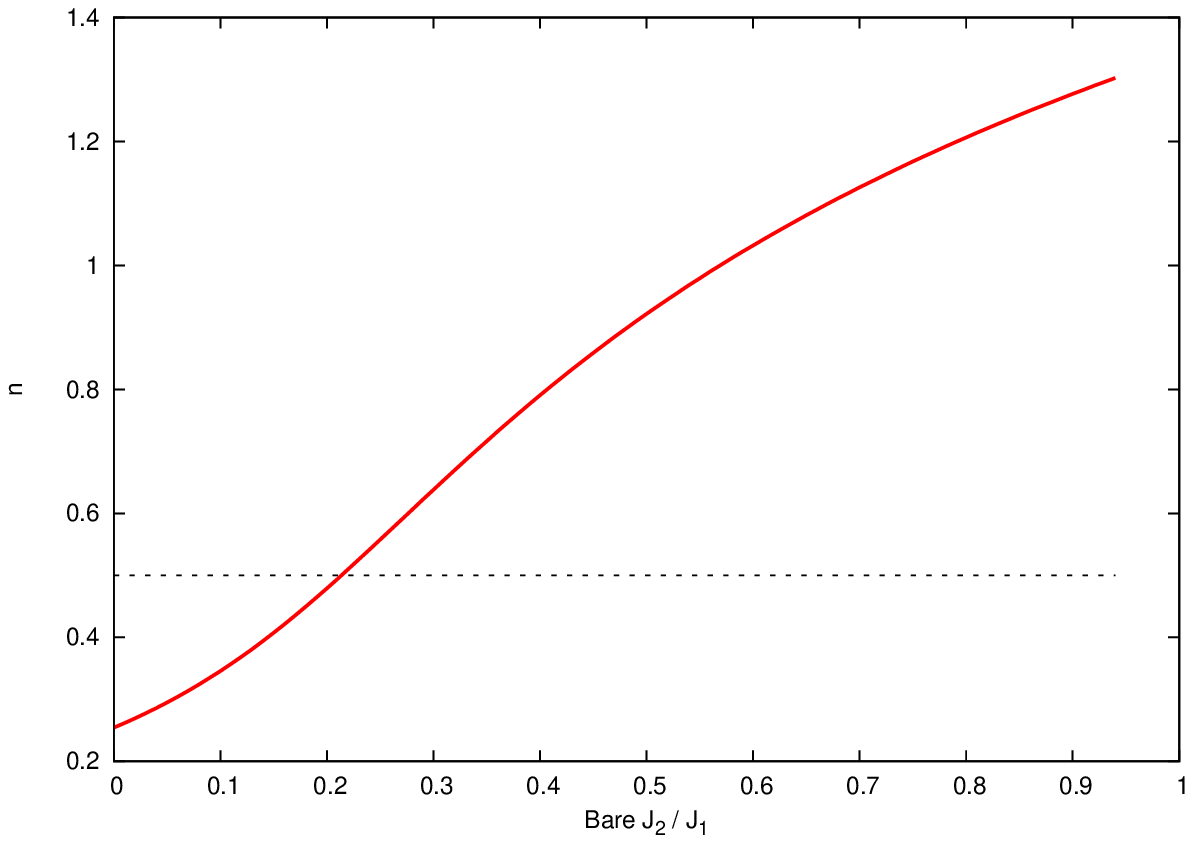}
\caption{Plot of the order parameter $n$. When $n\sim S$, N\'eel is expected to become unstable to quantum fluctuations. For $S=1/2$ (shown as a dotted line), this happens at $(J_2/J_1)_{bare} \sim 2.14$.}
\label{Fig.n}
\end{figure}

Clearly, the Hartree Fock decouplings merely renormalize the exchange couplings $J_1$ and $J_2$. We have
\bea
J_1 \rightarrow J_1 (1+\{p-n\}/S ); \phantom{abcd}
J_2 \rightarrow J_2 (1+ \{h-n\}/S ). \nonumber
\eea
We obtain the Hartree Fock parameters $n$, $h$ and $p$ self-consistently. For every `bare' value of $J_2/J_1$, we obtain a renormalized value of $J_2/J_1$ as plotted in Fig. \ref{Fig.effcoupling}. Fig.\ref{Fig.n} plots the obtained value of $n$. When $n\sim S$, the N\'eel moment is renormalized to zero and N\'eel order is expected to become unstable to quantum fluctuations. For $S=1/2$, this happens for $(J_2/J_1)_{bare} \sim 0.214$.

Our self-consistency equations are equivalent to a particular formulation of Schwinger Boson mean-field theory. This connection between two very different methods has been pointed out earlier for the case of the square lattice~\cite{Mila}. Thus, our value of critical $J_2/J_1$ is close to Schwinger Boson mean-field result of Ref.~\cite{Mattsson}. 

\subsection{Plaquette operators}

On a single plaquette, we denote the two Kekul\'e singlet covers as $\vert a \ra$ and $\vert b \ra$. We take these states to be normalized. They are however not orthogonal with $\la a \vert b \ra = -1/4$, upto a phase that can be gauged away. We denote symmetric (s-wave) and antisymmetric (f-wave) combinations of these covers as
\bea
\vert + \ra  = \sqrt{\frac{2}{3}} \left( \vert a \ra + \vert b \ra \right),
\vert + \ra  = \sqrt{\frac{2}{5}} \left( \vert a \ra - \vert b \ra \right).
\eea
With this definition, we have $\la + \vert + \ra = \la - \vert - \ra=1$ and $\la + \vert - \ra =0$. 
On an isolated plaquette, this operator takes the expectation values:
\bea
\la a \vert \hat{P} \vert a \ra = 1/2, \phantom{ab}\la b \vert \hat{P} \vert b \ra &=& 1/2\\
\la + \vert \hat{P} \vert + \ra &=& -3/4, \\
\la + \vert \hat{P} \vert - \ra &=& 0, \\
\la - \vert \hat{P} \vert - \ra &=& 5/4.
\eea

Our cluster
supports pRVB order. With pure pRVB ordering, the central plaquette must be in the $\vert - \ra$ state.
We find a positive expectation value for  $\hat{P}_{central}$ which supports the hypothesis that the central plaquette is in the $\vert - \ra$. However, the expectation value is less than 5/4 which indicates that quantum fluctuations reduce the strength of pRVB order.
Our cluster geometry can also accommodate c-VBS order, in which case the central plaquette would be in a pure $\vert a \ra$ or $\vert b\ra$ state. We have explicitly checked that this does not occur. 

\bibliographystyle{apsrev}
\bibliography{Heisenberg_honeycomb}